\RequirePackage{fix-cm}
\documentclass[smallextended]{svjour3}       
\smartqed  

\usepackage{appendix}
\usepackage{amsmath}
\usepackage{graphicx}
\usepackage[left]{lineno}
\usepackage{array}
\usepackage{longtable}
\usepackage{natbib}
\usepackage{booktabs}
\usepackage{ctable}
\usepackage{enumerate}
\usepackage{multirow}
\usepackage{hyperref}

\newcommand*\patchAmsMathEnvironmentForLineno[1]{%
\expandafter\let\csname old#1\expandafter\endcsname\csname #1\endcsname
\expandafter\let\csname oldend#1\expandafter\endcsname\csname end#1\endcsname
\renewenvironment{#1}%
{\linenomath\csname old#1\endcsname}%
{\csname oldend#1\endcsname\endlinenomath}}%
\newcommand*\patchBothAmsMathEnvironmentsForLineno[1]{%
\patchAmsMathEnvironmentForLineno{#1}%
\patchAmsMathEnvironmentForLineno{#1*}}%
\AtBeginDocument{%
\patchBothAmsMathEnvironmentsForLineno{equation}%
\patchBothAmsMathEnvironmentsForLineno{align}%
\patchBothAmsMathEnvironmentsForLineno{flalign}%
\patchBothAmsMathEnvironmentsForLineno{alignat}%
\patchBothAmsMathEnvironmentsForLineno{gather}%
\patchBothAmsMathEnvironmentsForLineno{multline}%
}

\ifpdf
    \graphicspath{{Figs/Raster/}{Chapter3/Figs/PDF/}{Chapter3/Figs/}}
\else
    \graphicspath{{Figs/Vector/}{Chapter3/Figs/}}
\fi

\begin{document}

\title{Urban boundary layers over dense and tall canopies
}


\author{Alexandros Makedonas         \and
        Matteo Carpentieri           \and
        Marco Placidi}


\institute{Alexandros Makedonas \and Matteo Carpentieri \and Marco Placidi* \at
              University of Surrey, Department of Mechanical Engineering Sciences, Environmental Flow Research Centre (EnFlo), Guildford, GU2 7XH, UK\\ 
              \\
              \email*{m.placidi@surrey.ac.uk}           
}

\date{Received: DD Month YEAR / Accepted: DD Month YEAR}

\maketitle

\begin{abstract}
Wind tunnel experiments were carried out on four urban morphologies: two tall canopies with uniform-height and two super-tall canopies with a large variation in element heights (where the maximum element height is more than double the average canopy height, $h_{max}$=2.5 $h_{avg}$). {The average canopy height and packing density were fixed across the surfaces to $h_{avg} = 80$ mm, and $\lambda_{p} = 0.44$, respectively.} A combination of laser doppler anemometry and direct drag measurements were used to calculate and scale the mean velocity profiles {within the boundary layer depth, $\delta$}. In the uniform-height experiment, the high packing density resulted in a `skimming flow' regime with very little flow penetration into the canopy. This led to a surprisingly shallow roughness sublayer ($z\approx1.15h_{avg}$), and a well-defined inertial sublayer above it. {In the heterogeneous-height canopies, despite the same packing density and average height, the flow features were significantly different.} {The height heterogeneity enhanced mixing thus encouraging deep flow penetration into the canopy. A deeper roughness sublayer was found to exist and extend up to just above the tallest element height (corresponding to  $z/h_{avg} = 2.85$)}, which was found to be the dominant lengthscale controlling the flow behaviour. {Results points toward the existence of an inertial sublayer for all surfaces considered  herein despite the severity of the surface roughness ($\delta/h_{avg} = 3 - 6.25$)}. This contrasts with previous literature.

\end{abstract}
\section{Introduction}
\label{intro}

Understanding flow around urban environments is becoming of increasing importance as cities and their populations grow in size \citep{desa2019world}. Although surface energy balance models have recently improved, accurate models for aerodynamic parameters are still poor, particularly for non-conventional roughness geometries - e.g. tall canopies with heterogeneous height, with urban flows that remain poorly described by both theoretical and empirical models \citep{kanda2013new}. The surface layer within the atmospheric boundary layer (ABL) is customarily split into two regions. Firstly, the roughness sublayer (RSL) which acts from the wall surface up to a certain point above the roughness elements. Customarily, the RSL is the region where the flow still encounters the effect of the individual roughness elements \citep{reading20193}. Secondly, the inertial sublayer (ISL), which covers a region above the RSL \citep{raupach1991rough}. The ISL begins at the top of the RSL and is characterised as a region of constant momentum flux \citep{reading20193}. Considerable debate has taken place over defining the boundaries of both the RSL and ISL. The height of the RSL has been commonly quoted between 2 - 5 times the average height ($h$) of the roughness surface \citep{raupach1991rough}, and more recently as low as 1.1 - 1.2 $h$ \citep{florens2013defining}. The ISL, and its existence, has been subject to even more debate. {Traditional theory \citep{stull2012introduction} suggests that in a sufficiently developed boundary layer an ISL should form. In contrast \cite{jimenez2004turbulent} postulated that for $\delta/h<80$ (where $\delta$ is the boundary-layer height), the RSL will increase in depth and effectively replace the ISL. \cite{jimenez2004turbulent} idea has been supported by several studies \citep{rotach1999influence, cheng2002near, cheng2007flow, hagishima2009aerodynamic}. However, \cite{leonardi2010channel} and \cite{cheng2007flow} argue that the relative boundary layer depth quoted by \cite{jimenez2004turbulent} may not be accurate.} 

Typically, in the standard characterisation of the effect of wall roughness, a surface (at least close to the wall) is characterised by the aerodynamic parameters and the friction velocity. It is of increasing importance to be able to determine the zero-plane displacement ($d$) and the roughness length ($z_{0}$) to effectively predict and calculate wind flow in and above the growing urban environments {(as discussed by \cite{stull2012introduction} in Sec. 9.7)}. In fully-rough conditions, these parameters are usually obtained, following \cite{cheng2002near}, by (i) calculating the friction velocity ($u_{*}$) and then (ii) applying logarithmic law fitting procedure in the constant flux layer (Eq. \ref{eq:U}).

\begin{align}
\hspace*{5mm} U=\frac{u_*}{k}\ln\left(\frac{z-d}{z_0}\right)
\label{eq:U}
\end{align}

The friction velocity can be determined in two ways. Directly, by calculating the drag generated by the rough wall e.g. using fully instrumented elements with static pressure ports or by floating element force balance \citep{cheng2007flow, hagishima2009aerodynamic, zaki2011aerodynamic}. These methods are both acceptable in the fully rough regime, where the viscous drag is almost negligible in comparison with form drag \citep{leonardi2010channel}. The second approach is indirect, where the friction velocity is evaluated from the Reynolds shear stress in the ISL \citep{cheng2002near, cheng2007flow}, so that $u_{*}=\sqrt{\tau_{0}~\rho^{-1}} \approx \sqrt{- \overline{u'w'}}$, where $\tau_{0}$ is the wall shear stress and ${\rho}$ is the density of the medium. {The roughness length $z_{0}$ is the height where wind speed reaches zero, and is used as a scale for the roughness of the surface \citep{stull2012introduction}}. The zero-plane displacement, $d$, is understood as a correction factor to the logarithmic profile \citep{cheng2002near,kanda2013new}, while physical reasoning tends to describe it as central height were drag occurs on a rough wall \citep{jackson1981displacement}. The aerodynamic parameters vary due to surface properties and, therefore, it should be possible to examine their impact on flow by systematically varying the surface characteristics. Many experiments have taken place over uniform-height cube arrays with different $h$, $\lambda_{p}$, and $\lambda_{f}$ (see \cite{grimmond1999aerodynamic} for $\lambda_{p}$ and $\lambda_{f}$ definitions) to examine how roughness affects the aerodynamic parameters \citep{cheng2002near, cheng2007flow, jackson1981displacement, grimmond1999aerodynamic, florens2013defining, macdonald1998improved, sharma2019turbulent}. However, the discrepancy in these studies with additional evidence from some recent numerical and experimental work, have demonstrated the inaccuracy of using just these variables to examine the sensitivity of the aerodynamic parameters \citep{hagishima2009aerodynamic, zaki2011aerodynamic, kanda2013new, reading20193, nakayama2011analysis, xie2008large}. Previous work has also highlighted how $\lambda_{p}$ or $\lambda_{f}$ - in isolation - are insufficient to characterise non-cubical roughness \citep{carpentieri2015influence}, and advocated for the need to decouple the two solidity ratios \citep{placidi2015effects,Placidi:2017}, however, this is outside the scope of this work. \cite{kanda2013new} pointed out the importance of two additional parameters when describing a canonical regular surface, the standard deviation ($\sigma_{h}$) and maximum element height ($h_{max}$). Others have attempted to determine the aerodynamic parameters for various environments by deriving semi-empirical relationships based on roughness geometry \citep{kanda2013new, macdonald1998improved, reading20193}. These have the advantage of being able to quickly determine the parameters without the need for field observations, wind tunnel tests, or computational experiments, allowing for fast prediction of flow in urban environments.

This article further explores how the aerodynamic parameters behave in extremely rough surfaces with heterogeneous heights in comparison with uniform-height cases at matching average height. The block arrays in this study are based on simplified attributes of super-tall grid cities, which feature a large standard deviation of element heights with large aspect ratios. A combination of variables, that to the authors' knowledge, have not yet been explored. The structure of this paper is as follows: the experimental set-up is discussed in Sec. \ref{sec:ExpSetup}. Sections \ref{sec:BLDepthI} - \ref{sec:ISDepthI} and \ref{sec:BLDepthII} - \ref{sec:ISDepthII} examine the depths of the boundary layer and surface layers. The aerodynamic parameters are then presented in Sections \ref{sec:ISLAveragedDataI} and \ref{sec:ISLAveragedDataII}. Finally, conclusions are drawn in Sec. \ref{sec:Conc}.

\section{Experimental facility and details}
\label{sec:ExpSetup}

\subsection{Experimental facility}

Experiments were conducted in the `Aero' tunnel within the EnFlo laboratory at the University of Surrey. This is a closed-circuit wind tunnel with a maximum speed of 40 m~s$^{-1}$. The free-stream velocity, measured by a Pitot tube located upstream of the model, was set to 10 m~s$^{-1}$ for all cases presented here. The tunnel's test section is 9 m long, 1.27 m wide, and 1.06 m tall. The streamwise, spanwise, and vertical directions are identified with the $x$, $y$, and $z$ axis, respectively. The $z$-axis is set from the top of the baseboard of the model (i.e. the actual wall), while the $y$ = 0 is set in the centre of the test section. The position $x$ = 0 is considered as the beginning of the tunnel test section. Time- and spanwise-averaged mean, and fluctuating velocities are denoted as ($U,V,W$), and ($u',v',w'$), respectively.  In the space between the beginning of the test section and the model a 1 m-long ramp rises from the floor of the tunnel to the average canopy height ($h_{avg}$ = 80 mm, Fig. \ref{fig:ArrayMap}). The ramp creates a smooth transition between the wind tunnel test section inlet and the roughness surface, thus minimising the flow disruption at the beginning of the roughness fetch \citep{cheng2002near} and allowing an equilibrium boundary layer to form. The most upstream measurement station is at $x$ = 3600 mm, where the flow is already fully developed. 
\subsection{Rough-wall models}
\label{sec:Models}

{Four surface roughnesses, all representing idealised tall and super-tall urban environments, were used in this study. Two of the surfaces have elements of uniform-height, while two have elements with varied-height. The individual roughness elements are sharp-edged cuboids of average height 80 mm. Urban buildings are classified by their aspect ratio, $AR=h/w$ (where $h$ and $w$ are the height and width of the building, respectively). Buildings with $3<AR<8$ fall in the tall regime (generally $100~m<h<300~m$), whilst buildings with $AR>8$ (and $h>300~m$) are referred to as super-tall \citep{CTBUHHeightCriteria}. Based on this criterion, the surface morphologies examined here are classified as tall or super-tall. Zero-pressure-gradient conditions were used in this work, as the acceleration parameter, calculated for both uniform- and varied- element height cases, was 4.85$\times10^{-8}$ and 9.48$\times10^{-8}$, respectively. The surfaces in examination are further described in the following sections.}

\subsubsection{Homogeneous-height model}\label{sec:HoHM}

To create an idealised urban environment in the case of uniform-height canopy, two urban features were studied: the packing density, {$\lambda_p$, and the element aspect ratio. Large urban areas with high-density buildings of 14 dense cities were used to calculate a characteristic urban packing density, as reported in Table \ref{table:CitPack}).} Google Maps was used to measure the area plots and building base sizes. A range between $\lambda_{p}$ = 0.33 to 0.57 was determined, which is in line with values cited for real cities in \cite{reading20193}. The packing density of $\lambda_{p}$ = 0.44 was within this range, and it was selected to describe densely packed cities. {The  \cite{CTBUH2015}} calculates a mean height of all buildings in Manhattan (New York) above 100 m to be 145.7 m. Using this average height and the bases of buildings from \cite{GMaps} gives an aspect ratio of approximately 3.4. Guided by this criterion, the effects of walls and fetch, a scaled model with elements of size 80 mm $\times$ 20 mm $\times$ 20 mm was designed. The roughness was mounted on five base plates. Each base plate when rotated allowed the roughness pattern to vary from aligned to 50 \% staggered, as in Fig. \ref{fig:Modules}. These surfaces are referred to as uniform-height aligned (UHA), and uniform-height staggered (UHS). In the homogeneous height roughness cases, a total of 5,775 elements were used.

\begin{table}[ht]
\caption{{Packing densities of highly dense areas within large urban areas. Values with an asterisk (*), are taken from \cite{grimmond1999aerodynamic}.}}
\def\arraystretch{1.3}
\centering
\label{table:CitPack}
\begin{tabular}{l l l l l l }
\toprule
City        & $\lambda_{p}$ &City        & $\lambda_{p}$    & City        & $\lambda_{p}$\\
\midrule

Beijing     & 0.50  & Los Angeles   & 0.36  & Singapore & 0.33 \\
Chicago     & 0.57  & Mexico City   & 0.47* & Tokyo     & 0.52 \\
Dubai       & 0.39  & Minneapolis   & 0.42  & Toronto   & 0.53\\
Hong Kong   & 0.53  & New York      & 0.57  & Vancouver & 0.39*\\
Kuala Lumpur& 0.35  & San Francisco & 0.49  &           &\\
London      & 0.43  & Shanghai      & 0.46  &           &\\

\bottomrule
\end{tabular}
\end{table}

\begin{figure}[ht!] 
\includegraphics[scale=0.425, trim={3cm 2cm 1cm 2cm},clip]{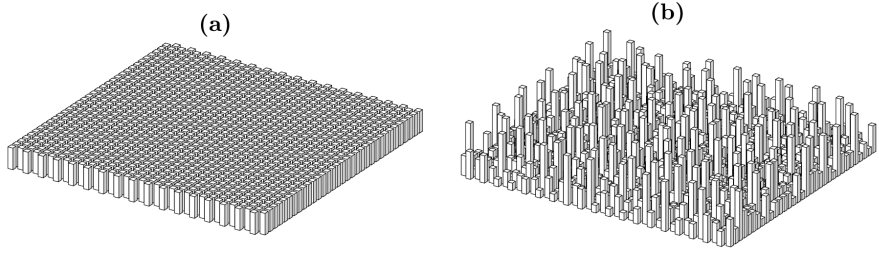}
\caption{{A single board of (a) homogeneous and (b) heterogeneous elements.}}
\label{fig:Planks}
\end{figure}

\begin{figure}
\centering    
\includegraphics[width=1.0\textwidth,height=0.3\textheight,keepaspectratio]{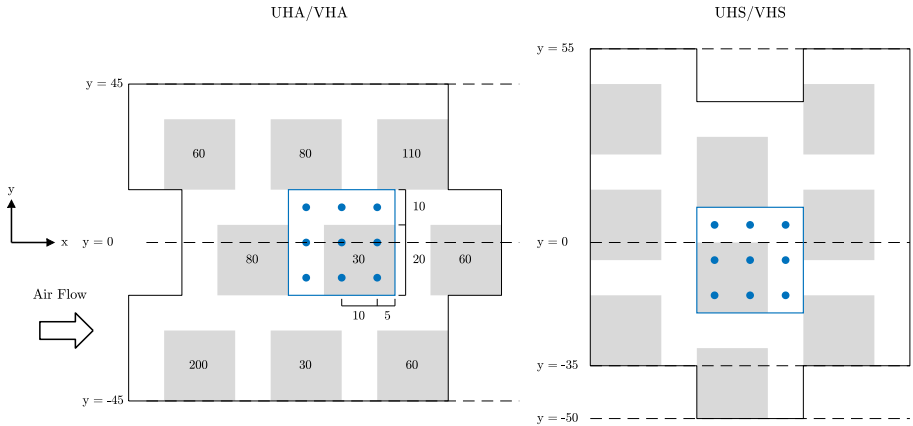}
\caption[]{{Aligned and staggered orientations of 3 $\times$ 3 module for homogeneous and heterogeneous models (all measurements are in mm). Indicated in the centre of the grey blocks are the heights of the elements in a heterogeneous module. Indicated with blue dots and black lines are an example of where the vertical profiles were measured within a repeating unit.}}
\label{fig:Modules}
\end{figure}

\subsubsection{Heterogeneous-height model}\label{sec:HeHM}

\begin{figure}
\centering    
\includegraphics[width=1.0\textwidth,height=0.3\textheight,keepaspectratio]{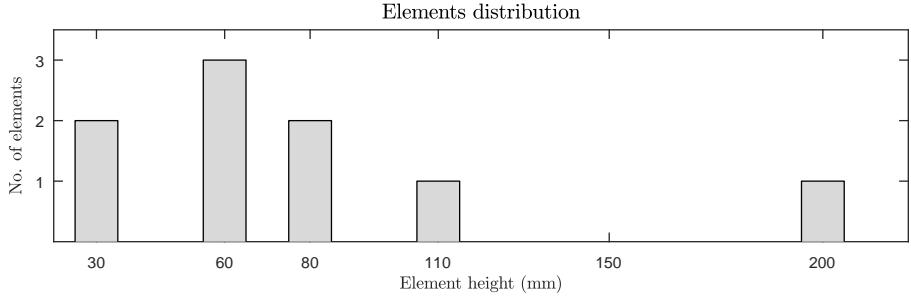}
\caption[]{{Probability density function of the distribution of element heights per module in heterogeneous model.}}
\label{fig:HHist}
\end{figure}

A varied-height model was also designed that differed from the uniform-height model in only one aspect, the standard deviation of element height ($\sigma_{h}$). The two different roughness configurations are shown in Fig. \ref{fig:Planks}, where the aligned homogeneous and heterogeneous surfaces are depicted in (a) and (b), respectively. The use of a large $h_{avg}$ provides the ability to introduce large $\sigma_{h}$. The standard deviation of the elements was modelled on the districts of Mong Kok (Hong Kong) and Midtown Manhattan. The geometric properties of the elements were derived from the \cite{NYCDoCP} and \cite{TownPB}, along with the buildings' design guide by \citet{doi:10.1680/mosd.41448}, when the information was lacking in the database. The real $\sigma_{h}$ of cities is 17.3 m and was scaled down to $\sigma_{h}$ = 0.049 m. As in the homogeneous model, the average height is 0.08 m, hereafter denoted as $h_{avg}$. Other than the $\sigma_{h}$ = 0.049 m, element heights were selected by matching the {distribution of building heights} in the data set, resulting in an increased number of short elements with only a few elements taller than the $h_{avg}$; $h_{max}$ being 0.2 m. The rough surfaces were constructed by assembling elements into modules. Herein a module is: 3 $\times$ 3 elements consisting of five different heights, randomly placed (Fig. \ref{fig:Modules}). In the heterogeneous model, a repeating module is needed to achieve statistically relevant statistics, as described by \cite{cheng2002near}. Each module contains elements with heights: 30, 30, 60, 60, 60, 80, 80, 110, and 200 mm, { distributed as in Fig. \ref{fig:HHist}}. The purpose of this randomisation is to {avoid creating preferential corridors for the flow between the super-tall elements facing the wind direction} and to create a more realistic city layout. The tallest elements in the varied-height surface are within the super-tall regime, with $AR=10$ \citep{jianlong2014study}. The varied-height surfaces have a {total of 5,445 elements.} Similarly to the homogeneous canopy, each base plate when rotated allowed for the roughness pattern to be modified from aligned to 50 \% staggered. Due to additional geometrical constrains, the total number of elements is slightly different for the heterogeneous and homogeneous canopies, {however, this is of little consequence as the downstream flow is spanwise self-similar over the different repeating units and fully developed in a central section of the wind tunnel (i.e. at the measurement stations).} The surfaces are referred to as varied-height aligned (VHA), and varied-height staggered (VHS). A summary of the surfaces can be found in Table \ref{table:Aligned and staggered orientations of models}.

\begin{table}[ht]
\caption{{Characteristics of the different roughness surfaces considered in this work.}}
\def\arraystretch{1.3}
\centering
\label{table:Aligned and staggered orientations of models}
\begin{tabular}{l c c c c c c}
\toprule
Case  &  \multirow{2}{5.5em}{Configuration} & \multirow{2}{3.5em}{Height} & $h_{avg}$ & $h_{max}$ & $\sigma_{h}$  & Elements\\
ID&&&(mm)&(mm)&(mm)&per module\\
\midrule
UHA & Aligned   & Uniform & 80 &  80 & 0   & 1 \\
UHS & Staggered & Uniform & 80 &  80 & 0   & 1 \\
VHA & Aligned   & Varied  & 80 & 200 & 49  & 9 \\
VHS & Staggered & Varied  & 80 & 200 & 49  & 9 \\
\bottomrule
\end{tabular}
\end{table}

\subsection{Instrumentation}\label{sec:Instu}

A two-component Laser Doppler Anemometer (LDA), FiberFlow from Dantec, was employed to measure two velocity components of the flow simultaneously. The laser beams were converged by a 300 mm focal length lens, which facilitated measurements in between elements. In the varied-height model, two elements of 200 mm height are located adjacent to each other in some configurations. In these situations, no measurements can be acquired lower than $z$ = 40 mm due to laser obstruction by the elements. The model was sprayed with black matt paint to minimise light reflections. The LDA probe was mounted onto a traverse system which could move the measurement volume in three-dimensions within the tunnel with sub-millimetre accuracy. An elliptical mirror was attached to the LDA probe to rotate one of the laser-beam couples by 90$^\circ$, effectively changing the component of velocity measured by the LDA.

Additionally, a fully instrumented pressure tapped element was used in the case of the uniform-height surfaces to measure the differential force across an element, hence its drag. To allow for these measurements one element was removed and replaced with an identical 3D-printed plastic element, fitted with a total of 25 static pressure ports on one of its faces. The element could then be rotated to allow for the differential pressure to be measured. The drag force was determined by integrating pressure distribution over the front and back of the pressure tapped element, allowing to measure the shear stress due to pressure ($u_{*}(p)$) as in \cite{cheng2002near}. The $u_{*}(p)$ was only determined for the uniform-height surface given the impossibility of replicating all random permutation of the different height elements for the heterogeneous roughness case. 
\subsection{Experimental details}

\begin{figure}[ht!] 
\centering    
\includegraphics[width=1.0\textwidth,height=0.3\textheight,keepaspectratio]{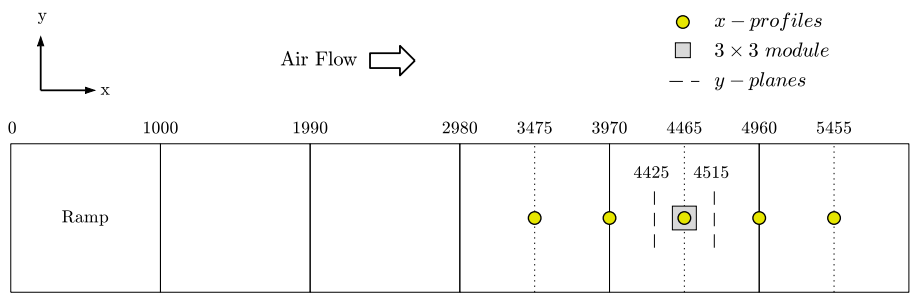}
\caption{{Schematic indicating the location of the LDA  measurements. All units are in mm.}}
\label{fig:ArrayMap}
\end{figure}

{LDA was used to measure velocity and Reynolds stress profiles within the ISL and RSL above the elements, but also within the canopy. For each surface, different types of LDA profiles were acquired as shown in Fig. \ref{fig:ArrayMap}. For both homogeneous and heterogeneous models \textit{x-profiles} were collected to study the development of the boundary layer. In the homogeneous model 9 vertical profiles were taken over a repeating element (see dots in Fig. \ref{fig:Modules}) to study the depths of the ISL and RSL. Over the varied-height surfaces 81 vertical profiles were taken over one module to study the statistical convergence of the Reynolds shear stress profiles within the RSL (i.e \textit{$3\times3$-module} in Fig. \ref{fig:ArrayMap}). Finally, over the heterogeneous model 18 vertical profiles across the width of the tunnel were taken to study the ISL depth (further details on \textit{y-planes} are contained in Fig. \ref{fig:Modules}).} 

\section{Results and discussion}\label{sec:RandD}
{This section presents and discusses results for the cases of uniform-height in \S \ref{sec:effofh} followed by results for the varied-height cases in \S \ref{sec:effofvh}.}
\subsection{Effect of uniform element height}\label{sec:effofh}

\subsubsection{Depth of the boundary layer}\label{sec:BLDepthI}
\begin{figure}[ht] 
\centering    
\includegraphics[width=1.0\textwidth,height=0.35\textheight,keepaspectratio]{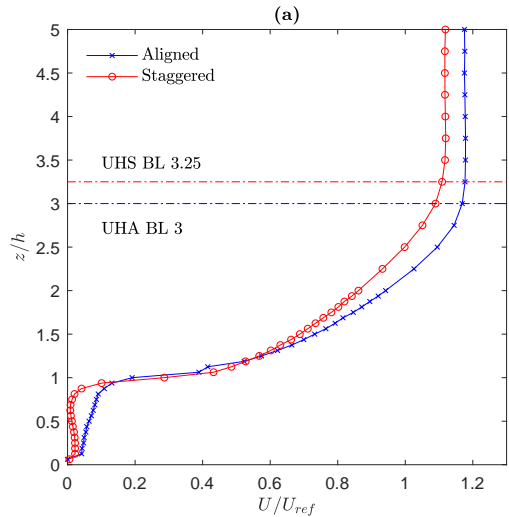}
\caption{Boundary layer depth shown with vertical velocity profiles over UHA and UHS models.}
\label{fig:BLUH}
\end{figure}

The height of the boundary layer, $\delta$, is commonly estimated as the height where $U$ = 0.99 $ U_{ref}$, where $U_{ref}$ is the freestream velocity. {The $U_{ref}$ used in all plots is measured at wind-tunnel inlet, ahead of the rough wall models. Since $z_{0}$ is a strong function of roughness fetch in a developing flow \citep{cheng2007flow}, the boundary layer must be in equilibrium with the surface below it and fully rough before representative measurements can be collected.} It was determined that this occurs over a fetch of $x$ = 4000 mm ($\approx 15$ $\delta$) for both varied- and uniform-height models. The results past this location are shown in Fig. \ref{fig:BLUH}, where $\delta$ is shown to be up to the height of 3.25 $h$ and 3 $h$ in the UHS and UHA cases, respectively. The increase in boundary layer thickness in the UHS case is likely due to increased frontal blockage of the staggered array. When the elements are staggered, `skimming' flow regime \citep{grimmond1999aerodynamic}, is less likely to occur, as the streamwise distances between the element increase, allowing for the development of the `wake interference' flow regime, with associated enhanced turbulent structures and an increased boundary layer thickness.

The current data ($h_{avg}$ = 80 mm and $\lambda_{p}$ = 0.44) is compared with the previous literature in Table \ref{table:AeroPar}. The taller elements in this experiment occupy a much larger fraction of the boundary layer ($\delta/h_{max}=3.25$) than those of \cite{cheng2002near} and \cite{cheng2007flow}, where $\delta/h_{max}=12$ and $\delta/h_{max}=7$, respectively; i.e. the surfaces described here have a higher relative roughness height. At the same freestream velocity, the elements of $h$ = 20 mm of \cite{cheng2002near} and \cite{cheng2007flow} show a much larger $\delta/h_{max}$ than the one found here, indicating that - not surprisingly - the height of the elements can influence the boundary layer depth. {For the sake of brevity, the reader is referred to \cite{andrieux2017} and \cite{thorpe2018} for a more in-depth discussion of the boundary layer variation across stations in both the streamwise and spanwise direction due to the surface heterogeneity.} 

\begin{table}[ht]
\caption{Surface characteristics of various roughness configurations. Bold cases are the current measurements. C20A, C20S, C10S and RM10S taken from \cite{cheng2002near} and C20A-25\%, C20S-25\%, C20A-6.25\% and C20S-6.25\% are from \cite{cheng2007flow}.} 
\def\arraystretch{1.3}
\centering
\label{table:AeroPar}
\begin{tabular}{l c c c c c c c c c}
\toprule
Case ID & $\lambda_{p}$ & $\delta/h$         & $RSL/h$   & $ISL/h$   & $u_{*}/U_{ref}$   & $d/h$   & $z_{0}/h$ & $Re_{\tau}$\\
   \midrule
\textbf{UHA}         & 0.44  &  3.00    & 1.13    & 1.63    & 0.072    & 1.12 & 0.007 & 1.14 $\times$ $10^{4}$\\
\textbf{UHS}         & 0.44  &  3.25    & 1.13    & 1.56    & 0.077    & 1.02 & 0.013 & 1.32 $\times$ $10^{4}$\\
\textbf{VHA}         & 0.44  &  6.25    & 2.85    & 4.75    & 0.094    & 2.55 & 0.043 & 3.10 $\times$ $10^{4}$\\
\textbf{VHS}         & 0.44  &  6.25    & 2.85    & 4.88    & 0.097    & 2.66 & 0.046 & 3.20 $\times$ $10^{4}$\\
   \midrule
C20A                & 0.25  &  7.55     & 1.85    & 0.55    & 0.061    & 1.18 & 0.023 & --\\
C20S                & 0.25  &  7.05     & 1.85    & 0.45    & 0.063    & 1.03 & 0.028 & --\\
C10S                & 0.25  &  12.1     & 1.80    & 1.40    & 0.058    & 1.16 & 0.012 & --\\
RM10S               & 0.25  &  13.7     & 2.50    & 0.80    & 0.063    & 1.36 & 0.014 & --\\
C20A-25\%           & 0.25  &  6.90     & 1.80    & 0.60    & 0.068    & 1.00 & 0.039 & --\\
C20S-25\%           & 0.25  &  6.70     & 1.75    & 0.40    & 0.071    & 0.96 & 0.045 & --\\
C20A-6.25\%         & 0.06  &  6.20     & 4.00    &   --    & 0.064    &--    & --    & --\\
C20S-6.25\%         & 0.06  &  6.80     & 1.80    & 0.40    & 0.072    & 0.62 & 0.044 & --\\
\bottomrule
\end{tabular}
\end{table}
\subsubsection{{Mean and fluctuating velocity profiles}}\label{sec:MF_profiles}
\begin{figure}[ht!] 
\centering    
\includegraphics[width=1.0\textwidth,height=0.7\textheight,keepaspectratio]{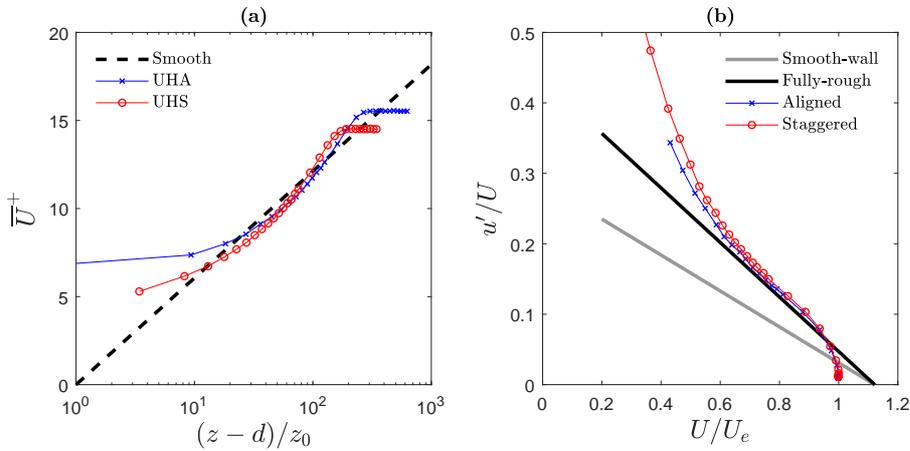}
\caption{{(a) Mean velocity profiles in inner scales for the UH cases. The dotted black line represents a smooth wall. (b) Diagnostic plot for the same cases. The black solid line represents the fully-rough regime \citep{Castro:2013}, while the grey line is the smooth-wall limit \citep{Alfredsson:2011}.}}
\label{fig:DP}
\end{figure}

{Mean streamwise velocity profiles for flow over uniform-height cases are shown in Fig. \ref{fig:DP}a. These profiles have been spatially averaged across one repeating unit to offer a fair representation of the surfaces under investigation and are taken in the region of fully developed flow. The units are non-dimensionalised with the skin friction velocity $u_{*}$ and the roughness length $z_0$. For both cases in Fig. \ref{fig:DP}a, a linear region within the profiles is evident; this offers strong support for the existence of a well defined ISL. Within this region, the collapse of the rough cases onto the smooth theoretical dashed line is an indication of the validity of the assumptions underpinning the methodology used to estimate the roughness parameters. This is further discussed in \S \ref{sec:ISLAveragedDataI}. The logarithmic region seems to extend well into what would be expected to be the roughness sublayer, as suggested in \cite{cheng2002near}. It must be stressed that data from different streamwise stations were considered for this analysis. The characteristics of the linear region visible in Fig. \ref{fig:DP}a, and the roughness parameters describing it, were found to be fully converged and independent from the streamwise location (i.e. from the boundary-layer depth) once the flow developed over approximately 15 boundary layer height. This offers a further proof that $15$ $\delta$ is the minimum required roughness fetch for the flow to be in equilibrium with the underlying surfaces considered herein. The parameters reported in Table \ref{table:AeroPar} are referring to these conditions, where the flow is fully developed. Next the streamwise turbulent fluctuations are discussed in Fig. \ref{fig:DP}b in the form of the diagnostic plot \citep{Alfredsson:2010}, which removes the need to define the roughness parameters and the friction scaling. The $U_{e}$ used to normalise the mean velocity in Figures \ref{fig:DP} and \ref{fig:DP2} is the local velocity at the edge of the boundary layer. The benefit of the diagnostic plot approach in this work is twofold, firstly it allows one to assess the universality and the self-similar character of the turbulent statistics, secondly it provides an independent assessment on whether the flow is in fully-rough conditions. For these reasons, the smooth \citep{Alfredsson:2011} and the fully-rough \citep{Castro:2013} asymptotes are reported in Fig. \ref{fig:DP}b (and Fig. \ref{fig:DP2}b). The data for uniform-heights (both for the aligned and staggered arrays) shows satisfying collapse onto the fully-rough line (black solid line). This is a strong indication that the turbulence statistics are self-similar across cases and that the boundary layer developing above the urban canopies in examination respects the classical scaling laws.}

\subsubsection{The roughness sublayer}\label{sec:RSDepthI}
\begin{figure}[ht!] 
\centering    
\includegraphics[width=1.0\textwidth,height=0.7\textheight,keepaspectratio]{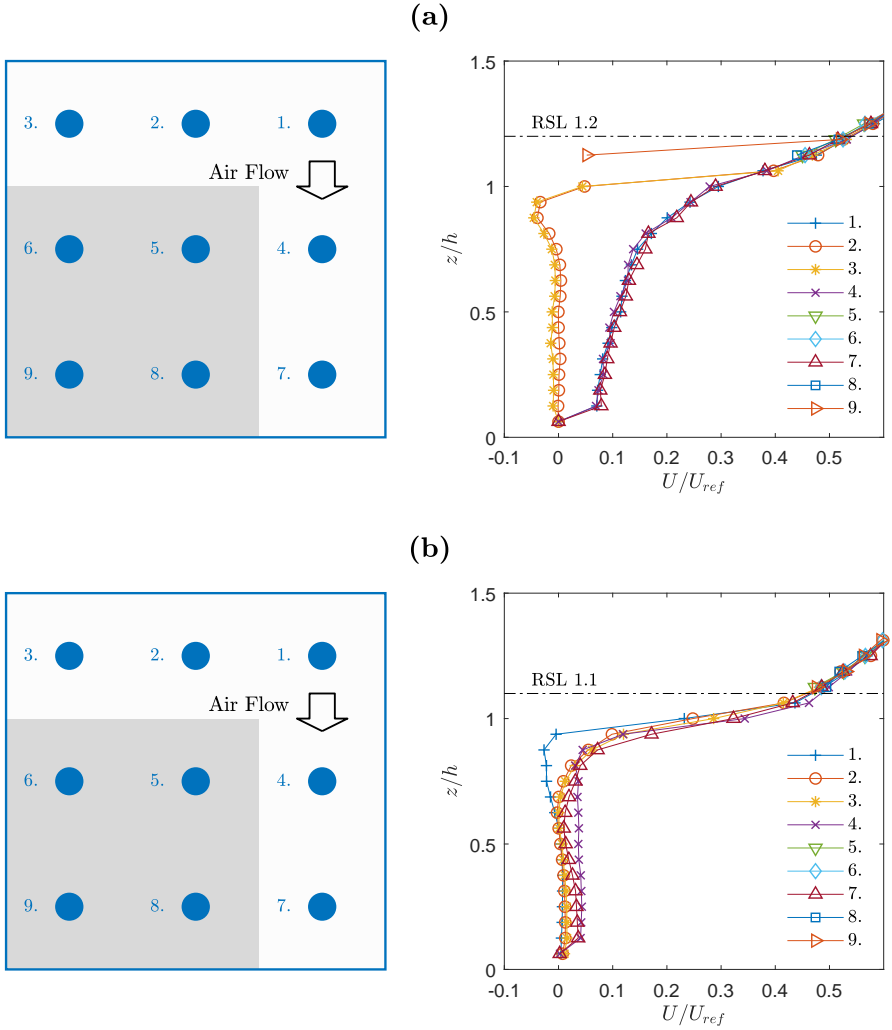}
\caption[]{On the right: RSL depth shown with vertical velocity profiles for uniform-height models. On the left: positions of profiles shown over a module: (a) UHA; (b) UHS. }
\label{fig:RSLUH}
\end{figure}

The top of the RSL is considered to be the point where all the effects of the individual roughness elements on the flow cease \citep{cheng2007flow, reading20193}. It follows that the flow inside the RSL is not spatially homogeneous. The nine vertical profiles taken over the surface are presented in Fig. \ref{fig:RSLUH}a,b for the UHA and UHS cases, respectively. These are shown to converge, unexpectedly, at about 1.15 $h_{avg}$ for both cases. The uniform-height surface results for this tall-element canopy differ from the small-cube cases analysed by \cite{cheng2002near}, where the RSL was found to be much deeper ($\approx 2$ $h_{avg}$). Despite the comparable Reynolds number, the RSL depth is much shallower here, closer to those found by \cite{florens2013defining}. Given the similar conditions between the current data and those of \cite{cheng2002near}, we attribute this difference to the discrepancy in the packing density and the much deeper canopy layer. The current results also vary significantly from the commonly cited 2 - 5 $h_{avg}$ \citep{reading20193,Flack:2007,roth2000review} due to the combination of high aspect ratio elements and dense canopy.

Furthermore, across Fig. \ref{fig:RSLUH}a and Fig. \ref{fig:RSLUH}b a clear difference can be seen between cases. Figure \ref{fig:RSLUH}b show similarity and collapse between all profiles. The flow inside the canopy behaves as predicted in literature, similar to that of a densely forested canopy, with a region of severe velocity deficit up to the elements' average height; {this is accordance with Fig. 9.7 in \S 9.7.3 from \cite{stull2012introduction}}. These results are qualitatively in agreement with those in \cite{Nept:2012b} for vegetating canopies for which an inflection point in the velocity profiles appears just above the roughness height for dense canopy (i.e. $\lambda_f>0.23$). The tightly packed roughness generates a strong shear layer at the top of the elements. \cite{cheng2007flow} argues that this region is the main contributor to $z_{0}$, as demonstrated in the 20 mm cube arrays. Figure \ref{fig:RSLUH}a, however, highlights the effect of aligned street canyons.  Profiles 1, 4, and 7 exhibit flow penetration into the canopy where the street canyons are aligned. The rest of the profiles follow a trend very similar to that in Fig. \ref{fig:RSLUH}b. This demonstrates that, for $\lambda_{p}$ = 0.44, the flow cannot penetrate deeply into the canopy, unless the streets are aligned. Even so, velocity profiles 1, 4, and 7 only begin to have a significant velocity increase at around 0.5 $h_{avg}$, indicating that at this $\lambda_{p}$, even in aligned street canyons, the penetration into the canopy is limited beyond the depth of 40 mm. A further study could be conducted on varied canopy depth models to verify how deep the flow can penetrate in aligned surfaces with varying $h$ at different packing densities. In summary, uniform tall arrays, concerning the RSL, behave similarly to forested canopies, showing an inflection point in correspondence with the top of the canopy. Lastly, for large $\lambda_{p}$ and $h/\delta$, even when the elements are aligned, the flow cannot penetrate significantly within the canopy.

\subsubsection{The inertial sublayer}\label{sec:ISDepthI}
\begin{figure}[ht!] 
\centering    
\includegraphics[width=1.0\textwidth,height=0.7\textheight,keepaspectratio]{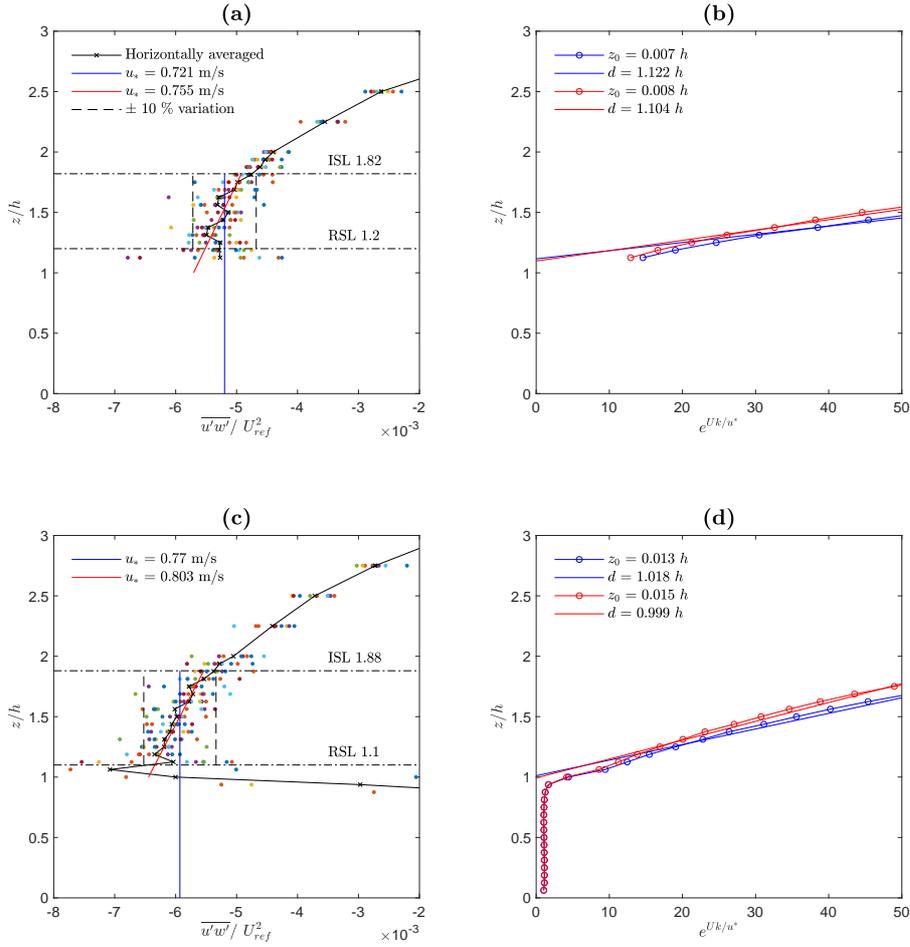}
\caption[]{{On the left: ISL shown with shear stress scatter plot and spatially averaged shear stress profile (black line). Black dotted lines show the ISL boundaries determined by the $\pm$ 10\% from the averaged ISL value. Blue line indicates $\overline{u'w'}$ determined by average ISL value, whilst the red line uses best-fit extrapolation to the $h_{avg}$ to determine $\overline{u'w'}$ (as in \cite{florens2013defining}). On the right: Eq. \ref{eq:U} is rearranged to the form of $z=z_0e^{kU/u_*}+d$, where $u_*$ is calculated using $\overline{u'w'}$. Using this rearranged form, $z_0$ represents the gradient and $d$ is the axis intercept. A least-mean-square fit is used to extrapolate onto the axis within the ISL boundaries. Case UHA for plots (a) and (b), and case UHS for plots (c) and (d).}}
\label{fig:ISLUH}
\end{figure}

In the ISL, the wall-normal variation of shear stress may be neglected, hence the constant-flux region denomination \citep{reading20193}. Here, we defined the ISL as the region where the vertical variation of the spatially averaged profiles of Reynolds shear stresses is below $\pm$ 10 \%, as in  \cite{cheng2007flow}. The profiles are reported in Fig. \ref{fig:ISLUH}a,c. The base of the ISL is assumed to be the top of the RSL found in Section \ref{sec:RSDepthI}. Unlike the work by \cite{cheng2002near} and \cite{cheng2007flow}, where the determination of an ISL was found to be challenging, here an ISL of significant depth can be observed in both cases. \cite{jimenez2004turbulent} reported $\delta/h\approx80$ as the lower limit for a canonical ISL to exist. For the uniform-height model, $\delta/h\approx3$, well below the limit where an ISL should form. Arguably, in Fig. \ref{fig:ISLUH} an ISL forms for these cases. \cite{cheng2007flow} also argued that the ISL depth is not constant which is also demonstrated here. As discussed in Section \ref{sec:BLDepthI}, it is likely that a deep ISL forms due to the large $\lambda_{p}$. {It is intuitive} to imagine that, in the limit of $\lambda_{p}\to\infty$, a new raised smooth wall surface (at $z=h$) will form, and recover the canonical turbulent boundary layer structure. {Additionally, it was discussed in \S \ref{sec:MF_profiles} how both the mean velocity profiles and the turbulent statistics were found to conform to the canonical scaling for turbulent boundary layers, and how the roughness parameters were invariant with streamwise location once the flow was fully-developed. These findings strengthen the argument that, despite the relative roughness height considered in this work, an ISL does indeed develop over these urban surfaces.}

\subsubsection{The aerodynamic parameters}\label{sec:ISLAveragedDataI}

The method used in this study to calculate $d$ and $z_{0}$ uses a logarithmic law fitting. A common problem faced with the uncertainty in the fitting of the log-law is that there are three free parameters, $u_{*}$, $d$ and $z_{0}$ \citep{Castro:2007,Segalini:2013}. More generally, other unknowns are given by the wake parameter, $\Pi$, and by not considering the Von K\'{a}rm\'{a}n `constant' a constant \citep{castro2017measurements}, however, this is not relevant for the current discussion. To reduce the uncertainty in the fitting procedure, a common method involves first fixing $u_{*}$. This is done here with two methods: (i) by using the Reynolds shear stress value in the ISL to determine $u_{*}$, and (ii) by measuring the drag directly by use of pressure-tapped elements. In the first method, the horizontally-averaged vertical velocity profile is used in either of two ways. Firstly, one can simply compute the average value from all the points within the ISL; secondly, a best-fit extrapolation onto the height z = $d$ or z = $h$ can be used \citep{cheng2002near, cheng2007flow, florens2013defining}. An independent method to compute $u_{*}$ is with the use of a pressure instrumented element, as described in \cite{cheng2002near}. Results are summarised in Table \ref{table:AeroPar}. Once the friction velocity has been determined, the aerodynamic parameters are computed by best fitting the velocity profiles considering the Von K\'{a}rm\'{a}n's constant $\kappa = 0.4$, which is within the limits suggested by \cite{Marusic:2013} for high Reynolds number flows. {The discrepancy between} the friction velocities calculated via averaged and extrapolated ISL is below 4 \% (Fig. \ref{fig:ISLUH}b,d). A further comparison extrapolating to z = $h$ \citep{florens2013defining} gives results more closely resembling the pressure tapped results (i.e. $\approx$ 1.5 \%  and $\approx$ 15 \% for $d$ and $z_{0}$ respectively), suggesting that the extrapolation technique does yield to more robust results.
\subsection{Effect of varying element height}\label{sec:effofvh}


\subsubsection{Depth of the boundary layer}
\label{sec:BLDepthII}

\begin{figure}[ht] 
\centering    
\includegraphics[width=1.0\textwidth,height=0.3\textheight,keepaspectratio]{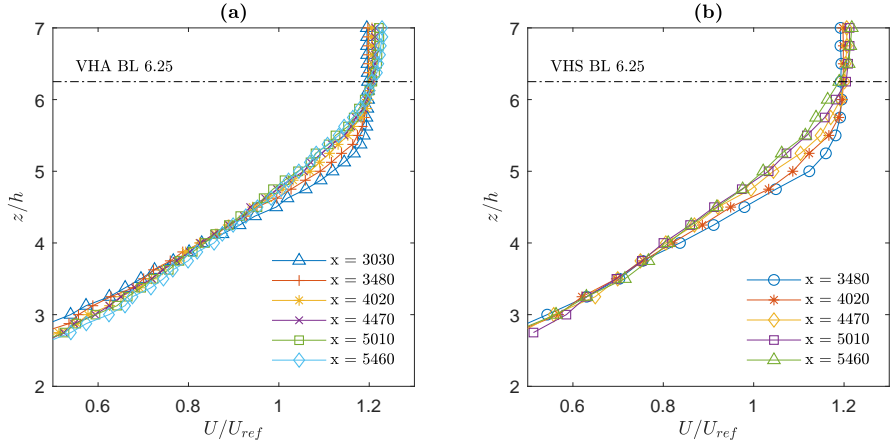}
\caption[]{Boundary layer depth found from vertical velocity profiles taken at different fetch over models: (a) VHA, and (b) VHS.}
\label{fig:BLVH}
\end{figure}

In the varied-height models, similarly to the previous case, an initial investigation was carried out to determine the fetch at which the boundary layer is fully {developed} and this was found to be at $x$ = 4470 mm ($\approx9$ $\delta$), as in Fig. \ref{fig:BLVH}. Although the $h_{avg}$ is the same in both models, the boundary layer depth is approximately double in the varied-height model ($\delta_{UHA}/h_{avg} = 3$ versus $\delta_{VHA}/h_{avg} = 6.25$). It is likely that the increased $\sigma_{h}$ also increases the drag generated by the surface, hence creating deeper boundary layers. 

\subsubsection{{Mean and fluctuating velocity profiles}}\label{sec:DP_2}
\begin{figure}[ht!] 
\centering    
\includegraphics[width=1.0\textwidth,height=0.7\textheight,keepaspectratio]{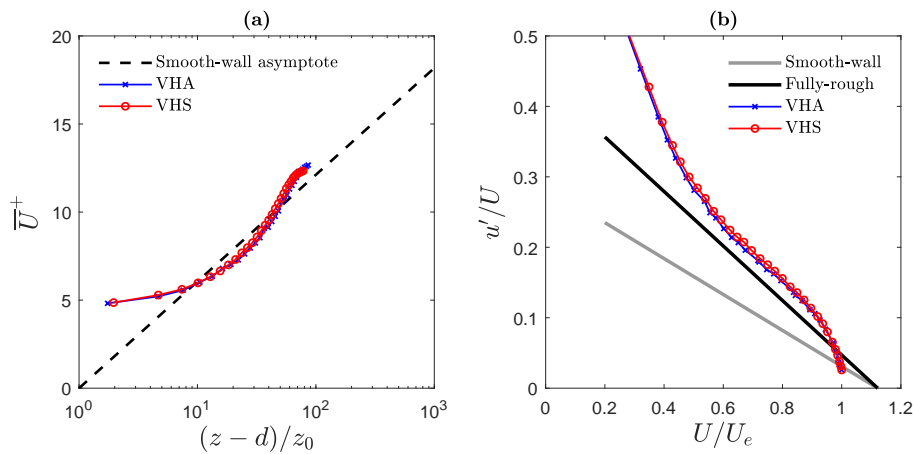}
\caption{{(a) Mean velocity profiles in inner scales for the VH cases. The dotted black line represents a smooth wall. (b) Diagnostic plot for the same cases. The black solid line represents the fully-rough regime \citep{Castro:2013}, while the grey line is the smooth-wall limit \citep{Alfredsson:2011}.}}
\label{fig:DP2}
\end{figure}

{As for the case of uniform-height previously discussed, mean and fluctuating velocity profiles are considered next. Mean streamwise velocity profiles for flow over varied-height cases are shown in Fig. \ref{fig:DP2}a. As for the previous case, a region of near-linear behaviour is noticeable, however, both the degree of collapse onto the smooth wall dashed line and the extent of the linear regime are reduced from the previous cases in Fig. \ref{fig:DP}a. Both these aspects are discussed in the following sections, however, these seem to indicate that the methodology for the evaluation of the roughness parameters (see \S \ref{sec:ISDepthII}) has a higher uncertainty for surfaces with height heterogeneity and that the roughness sublayer has grown at the expense of the inertial sublayer (see \S \ref{sec:RSDepthII}), respectively. It is still important to highlight that, even for this cases, data from different streamwise stations yielded converged roughness parameters; a strong evidence of the fully-developed nature of the flow at the measuring stations.}

{The diagnostic plot for the streamwise fluctuations is shown in Fig. \ref{fig:DP2}b. Both cases for varied-heights are found to be self-similar, however interestingly, these sit above the fully-rough trend line. This can be interpreted as an indication of the enhanced turbulent mixing due to the heterogeneous surface morphology; an aspect further discussed in \S \ref{sec:RSDepthII}. This discrepancy with previous data on fully-rough walls reported in \cite{Castro:2013} can also be attributed to the possibility that this fully-rough asymptote within the diagnostic plot is effected by several other roughness parameters, as highlighted originally in \cite{Castro:2013}. Two parameters of particular relevance to this work are the much higher relative roughness height, $\delta/h$, and the standard deviation in elements height, $\sigma_h$, which distinguish this work from that data in \cite{Castro:2013}.}
\subsubsection{The roughness sublayer}
\label{sec:RSDepthII}

\begin{figure}[ht!] 
\centering    
\includegraphics[width=1.0\textwidth,height=0.7\textheight,keepaspectratio]{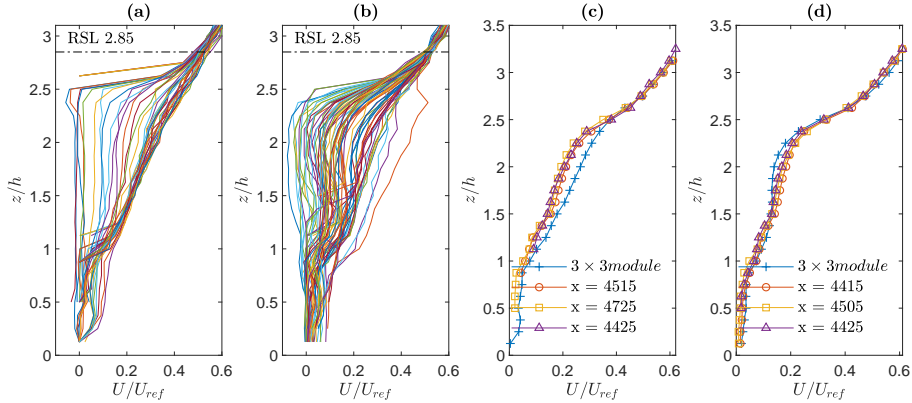}
\caption[]{RSL shown with vertical velocity profiles over the varied-height models. (a) VHA and (b) VHS. Horizontally-averaged vertical velocity profiles of different experimental runs. (c) VHA (d) VHS.}
\label{fig:RSLVH}
\end{figure}

In the varied-height experiment, 81 profiles were taken per repeating $3\times3$-module (Fig. \ref{fig:Modules}). These are shown in Fig. \ref{fig:RSLVH}a, b for the two configurations. There is a clear collapse of mean vertical profiles, indicating the existence of a limited RSL depth (Fig. \ref{fig:RSLVH}a,b). The RSL height is the same in both VHA and VHS configurations, which is of interest; it seems that the dominant feature in determining the height of the RSL is the tallest element in the $3\times3$-module, as a collapse is found for $z/h_{avg}\approx2.5$ ($z/h_{max}\approx1$). The evident collapse of vertical profiles in the RSL contradicts \cite{cheng2002near}, \cite{cheng2007flow} and \cite{jimenez2004turbulent}, who predicted an RSL region that expands and `squeezes' the ISL for rough walls with significantly tall and heterogeneous elements. For the VHA surface (Fig. \ref{fig:RSLVH}a) not all profiles collapse at the same height. Several vertical profiles converge near the average height of the elements (i.e. $z/h_{avg}$ = 1). The remaining profiles converge just above the maximum-height element ($z/h_{avg}$ = 2.5 or $z/h_{max}\approx1$). This trend possibly occurs for the same reasons discussed for the uniform-height cases. Where street canyons are aligned in the wind direction, the flow penetrates deeper into the canopy, allowing for higher velocities and enhanced mixing. Likewise, in the VHS configuration (Fig. \ref{fig:RSLVH}b), the flow does not seem to fully converge until above the tallest element ($z/h_{avg}$ = 2.85). In the varied-height situation skimming no longer occurs since there is a large spread of velocities below the $h_{avg}$ and $h_{max}$. As seen in both the VHA and VHS cases, the large spread of velocities within the canopy indicates significant flow penetration likely due to the increased $\sigma_{h}$. Considering this possibility, cities designed with large $\sigma_{h}$, could introduce much higher rates of mixing, with positive outcomes on urban air quality, and natural ventilation. As argued by \cite{zaki2011aerodynamic}, $\lambda_{p}$ is no longer a good indicator of the flow regime when a large $\sigma_{h}$ is introduced, which is counter to the studies conducted by \cite{sharma2019turbulent}. The argument made here suggests that the surface parameters used to describe flow over uniform-height models are no longer sufficient to characterise canopies with significant height variations. 

Figures \ref{fig:RSLVH}c and d show horizontally-averaged profiles within the RSL for the different experimental runs in the varied-height model. The spatially-averaged profiles taken in different areas of the array (across the wind tunnel span) collapse reasonably well onto each other, but there are some discrepancies around the region $1.5<z/h<2.5$. The profiles measured across the $y$-planes collapse better than the profiles taken over the $3\times3$-module. Possibly, it is more efficient and easier to get a representative averaged profile from taking measurements across the width of the model rather than over a single repeating module. Another reason for this better collapse could be that the average height of the elements in-front and behind the $y$-planes did not always match $h_{avg}$- see further discussion on this topic in \cite{cheng2002near}. However, the clear collapse in $y$-plane profiles indicates that $h_{avg}$ may not be the dominant lengthscale in models with large $\sigma_{h}$. Comparing uniform- to varied-height canopies across Fig. \ref{fig:RSLUH} and Fig. \ref{fig:RSLVH}, a prominent difference appears. For the varied-height models, the wind velocities vary greatly within the canopy, in contrast, they remain close to zero until the very top of the canopy for the uniform-height experiments. This is likely due to the physics of the `skimming flow' regime \citep{grimmond1999aerodynamic}. Even below $h_{avg}$, there are much higher velocities in the varied-height canopy compared to the uniform-height. The spread of velocity profiles in the varied-height model suggests that reasonable mixing can occur in the near-wall region for tall and dense canopy (large $\lambda_{p}$ and $h_{avg}$), providing that $\sigma_{h}$ is significant.

\subsubsection{The inertial sublayer}\label{sec:ISDepthII}
Several thresholds have been used in the literature for defining this region, where $\pm$ 5, $\pm$ 10, and $\pm$ 20 \% variation are all examined by \cite{cheng2007flow}. \cite{kanda2013new} used a region for logarithmic law fitting which is a function of the tallest and average roughness elements' height. \cite{sharma2019turbulent} noted how the ISL depth in their work is a function of the spacing between the roughness elements. These ISL fitting methods, however controversial, can prove very useful in accurately estimating aerodynamic parameters. However, a convincing relationship between the roughness geometry and the upper limits of the ISL is still elusive, particularly for tall canopies.

\begin{figure}[ht] 
\centering    
\includegraphics[width=1.0\textwidth,height=0.65\textheight,keepaspectratio]{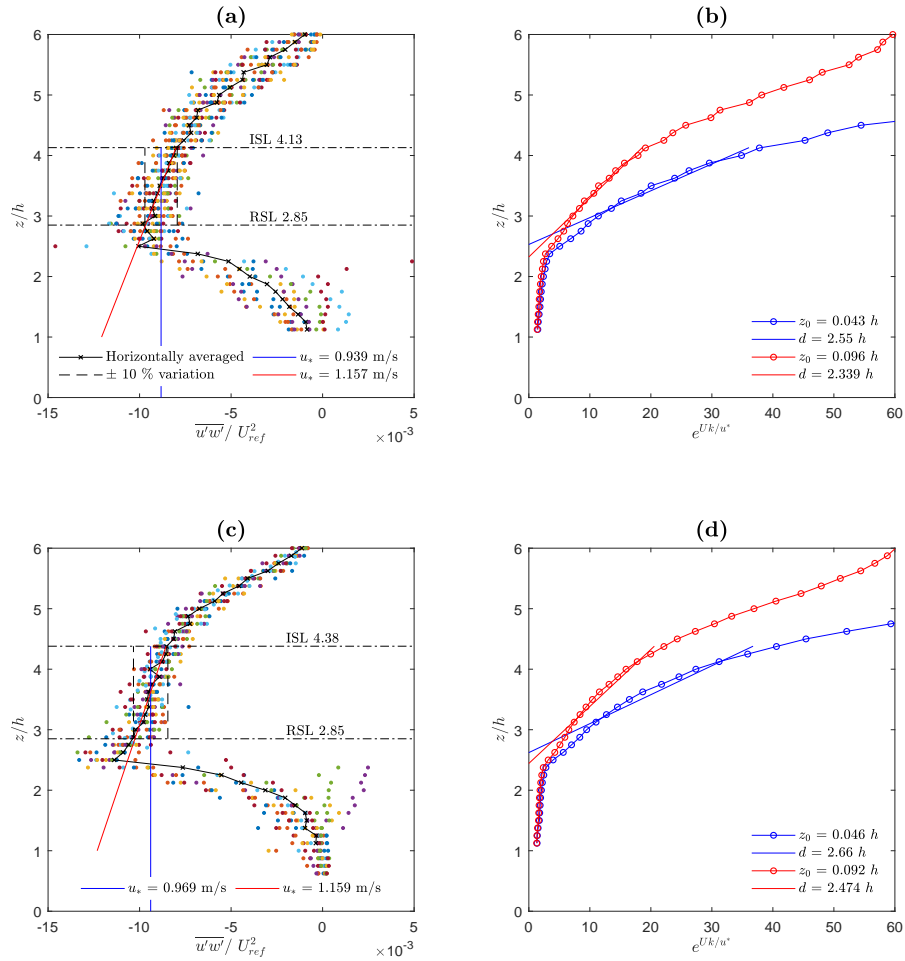}
\caption[]{{On the left: ISL shown with shear stress scatter plot and spatially averaged shear stress profile (black line). Black dotted lines show the ISL boundaries determined by the $\pm$ 10\% from the averaged ISL value. Blue line indicates $\overline{u'w'}$ determined by average ISL value, whilst the red line uses best-fit extrapolation to the $h_{avg}$ to determine $\overline{u'w'}$ (as in \cite{florens2013defining}). On the right: Eq. \ref{eq:U} is rearranged to the form of $z=z_0e^{kU/u_*}+d$, where $u_*$ is calculated using $\overline{u'w'}$. Using this rearranged form, $z_0$ represents the gradient and $d$ is the axis intercept. A least-mean-square fit is used to extrapolate onto the axis within the ISL boundaries. Case VHA for plots (a) and (b), and case VHS for plots (c) and (d).}}
\label{fig:ISLVH}
\end{figure}

The depth of the ISL region, herein based on \cite{reading20193}, is shown in Fig. \ref{fig:ISLVH}. Previous studies predicted that an ISL region would vanish as the RSL rises through the boundary layer due to increasingly rough surfaces \citep{cheng2007flow, cheng2002near, rotach1999influence, jimenez2004turbulent, hagishima2009aerodynamic}. The clear collapse of the RSL in Fig. \ref{fig:RSLVH} contradicts this theory. Furthermore, the $\pm$ 10 \% variation definition does allow for an ISL with significant depth to be singled out, as seen in Fig. \ref{fig:ISLVH}. This may be an indication that the ISL does, indeed, exist. It is important to point out, however, that the Reynolds shear stresses never reach a full plateau, but show a small - yet visible - gradient across the ISL. This is possibly related to the fact that the boundary layer in the cases with varied-height occupies a significant height of the wind tunnel. The acceleration parameter is fairly small (9.48$\times10^{-8}$) indicating a zero-pressure-gradient, however, it must be acknowledged that it is nearly double the same quantity in the uniform-height case. An ISL within the $\pm$ 10 \% variation does form even though $\delta/h= 6.25$ is still well below the limits established by \cite{jimenez2004turbulent}. \cite{florens2013defining} argued that this ratio should be lowered, and additionally \cite{cheng2007flow} demonstrates that the ratio $\delta/h$ should not be the only criterion for the existence of the ISL. Furthermore, \cite{leonardi2010channel} argued that the Reynolds number previously thought to be necessary for the ISL development may be much lower than expected. The finding from the current work, therefore, offers support to the idea that \citeauthor{jimenez2004turbulent}'s (\citeyear{jimenez2004turbulent}) ratio is perhaps too stringent. The $\pm$ 10 \% horizontal variation may not, however, be the best suit for determining the ISL depth. The top limit of the ISL can be hard to determine, and it is here tempting to soften the criterion to include the region where the gradient of the slope is more noticeable (i.e. $2.5<z/h<5$ in Fig. \ref{fig:ISLVH}c). Indeed, in the uniform-height experiment discussed earlier, the $\pm$ 10 \% horizontal variation is quite liberal, and a more constrained $\pm$ 5 \% could be used. The depth of the ISL could perhaps be better determined by the change in slope in the spatially-averaged Reynolds shear stresses, in a procedure similar to that in \cite{kanda2013new}. 
The relevance of this discussion stems from the fact that, if sufficiently accurate $d$ and $z_{0}$ estimates can be derived using the constant flux approximation, then wind-tunnel experiments could be greatly simplified, omitting pressure tapped elements or floating force balances, without compromising on the accuracy of the inner scaling. A summary of the depths of ISL and RSL for both cases of heterogeneous heights can be seen in Table \ref{table:AeroPar}. The heights of RSL, ISL, and boundary layer seem to be strongly dependent on the height of the tallest element within the surfaces. Even with appropriate normalisation (e.g. by $h_{avg}$ or $h_{max}$), there is no apparent universal scaling between results from varied- and uniform-height experiments. The significant change in the boundary layer, RSL, and ISL depths across the two cases seem to be uncorrelated. Thus, VHA and VHS surfaces could only be compared to models of similarly sized standard deviation and average height.

\subsubsection{The aerodynamic parameters}\label{sec:ISLAveragedDataII}
The aerodynamic parameters and friction scaling were determined as in Section \ref{sec:ISLAveragedDataI}, but direct measurement of the friction velocity by an instrumented element were not available for these cases for reasons discussed in Section \ref{sec:Instu}. Here we follow the same procedure explained in Section \ref{sec:ISLAveragedDataI}, i.e. {extrapolating to $z = h$}. The results are reported in Fig. \ref{fig:ISLVH}. The aerodynamic parameters calculated herein are compared to the predictions of \citeauthor{kanda2013new}'s (\citeyear{kanda2013new}) and \citeauthor{MacDonald:1998}'s (\citeyear{MacDonald:1998}) morphometric methods which considered elements with varied-height, to assess their performance. There is a large discrepancy between the varied-and uniform-height results, where $d$ is found to increase 1.5 times from uniform- to varied-height, whilst $z_{0}$ increases nearly 3.5 times. For the uniform canopy, \citeauthor{kanda2013new}'s (\citeyear{kanda2013new}) method predicted reasonable $d$ while \citeauthor{MacDonald:1998}'s (\citeyear{MacDonald:1998}) performed worse ($\approx$ 5 $\%$ and 30 $\%$, respectively). Neither produced a value for normalised $z_0$ that was close to that obtained herein (with both methods predicting a roughness length almost an order of magnitude larger). For the varied-height experiment, the extrapolated results from the shear stress plot produced a value of $d$ above $h_{max}$, whilst the average shear stress method resulted in values below $h_{max}$ (see Fig. \ref{fig:ISLVH}). Both \cite{kanda2013new} and \cite{Hopkins:2012} methods predict a value of $d$ between $h$ and $h_{max}$, which differ $\approx$ 36 $\%$ and $\approx$ 43 $\%$, respectively from those extrapolated here from the average shear. This discrepancy may be due to the fact that Tokyo (heavily used in \cite{kanda2013new}) does not have many super-tall structures such as Hong Kong (largely due to the seismic restrictions in the former). The lack of similarity of results continues when comparing $z_{0}$. The $z_{0}$ values calculated with the two methods were roughly twice as large those evaluated here.

The findings of this work strengthen the argument by \cite{kanda2013new} that new parameters are necessary to accurately characterise an urban rough wall, particularly when the elements are tall and heterogeneous in height if one wants to accurately estimate the aerodynamic parameters. \cite{jiang2008systematic} and \cite{zaki2011aerodynamic} suggest that $d$ increases with $\sigma_{h}$, which is in line with findings presented in this work. \cite{zaki2011aerodynamic} also demonstrated that the values of $d$ and $z_{0}$ almost doubles for cases with the same average height but significant $\sigma_{h}$ variation. \cite{xie2008large} also acknowledges that the tallest elements have a disproportionate contribution to the drag across a surface. \cite{kanda2013new} further suggested that both $\lambda_{p}$ and $\sigma_{h}$ have an effect on $d$ and $z_{0}$. A lack of similarity between results of uniform- and varying-height strengthens the above arguments; particularly the fact that a large $\sigma_{h}$ produces significant changes in the flow field. \cite{kanda2013new} considered $h_{max}$ as the upper limit of $d$, and also advocate the use of standard deviation of element height ($\sigma_{h}$) as an important measure when calculating the aerodynamic parameters, which is corroborated by the findings of this work. {This seems to support the argument that} the sole use of $\sigma_{h}$ and $h_{avg}$ when designing an urban model are not representative of the flow physics in a real city as suggested by \cite{kanda2013new}. Since there are usually a few very tall elements amongst many shorter elements, other important parameters to consider would be the distribution and ratios of tall to short elements.

\section{Conclusions}
\label{sec:Conc}

Wind tunnel experiments were conducted at the University of Surrey on four dense ($\lambda_p=0.44$) and tall ($\delta/h_{avg}\approx3$) urban arrays; two canopies were of uniform-height and two of varied-height, whilst the average height was kept fixed at $h_{avg}$ = 80 mm in both cases. All canopies aimed to represent idealised modern cities. The experiments examined the differences in the flow features across canopies with homogeneous and heterogeneous heights by means of laser doppler anemometry and direct drag measurements. All cases examined were fully-developed, fully-rough surfaces in zero-pressure-gradients.

In the uniform-height experiments, the surfaces with a large $\lambda_{p}$ inhibited deep penetration of the wind into the canopy, thus hindering mixing at street level. For $\lambda_{p}$ = 0.44, `skimming flow' regime seemingly occurred, and the rough surface started to recover smooth-like properties above the elements, producing an inertial sublayer region; this is in contrast to \cite{cheng2002near} and \cite{cheng2007flow} who questioned the existence of this region as the roughness influence grows. {The RSL depth was found to extend approximately up to $z$ =  1.15 $h$, which is much shallower than the typically expected 2 - 5 $h$ \citep{Flack:2007}. Conventional turbulent scalings laws were found to be applicable to both mean velocity profiles and turbulent quantities despite the severity of the roughness.} 

For heterogeneous-height canopies, the usefulness of $\lambda_{p}$ in describing the wall properties became more questionable. Despite the same $h_{avg}$, the boundary layer grew to almost double the thickness of the cases with uniform-height, highlighting the significance of $\sigma_{h}$ and $h_{max}$ in dictating the flow features. Surprisingly, even with such a heterogeneous canopy, a clear collapse of vertical profiles of Reynolds shear stress was observed, forming a coherent RSL extending to just above the height of the tallest element ($z/h_{avg}=2.85$ and $z/h_{max}\approx1$), {which is much shallower than anticipated. Moreover, more significant wind penetration was observed within the canopy when compared with the uniform-height array. This is responsible for an enhanced turbulent mixing resulting in velocity fluctuations which are higher than previously reported in surfaces with homogeneous heights throughout the boundary layer depth \citep{Castro:2013}.} These findings strengthen the need to include information regarding $\sigma_{h}$ and $h_{max}$ when describing flow over tall and heterogeneous canopies, supporting the conclusions highlighted in \cite{kanda2013new}. A comparison of the aerodynamic parameters for the cases with heterogeneous height considered herein with existing morphometric methods has highlighted the inaccuracies of these in the case of tall canopies with significant variation in height between elements. Further work is therefore needed on this topic.

\begin{acknowledgements}
{The authors would like to acknowledge Dr. Paul Hayden at the University of Surrey for facilitating the tests and the Department of Mechanical Engineering Sciences for funding the manufacture of the experimental rig. We are also grateful to Jacques Andrieux, Harry Thorpe, and Amal Pawa, who manufactured the UH model during their undergraduate projects and carried out preliminary tests.  Finally, we would like to acknowledge the support of the IMechE, via the Conference Travel Grant that allowed us to disseminate some of this work at the American Meteorological Society in 2020 (\url{https://ams.confex.com/ams/2020Annual/meetingapp.cgi/Paper/370772}).} 
The data used in this work is available at the following link: \url{https://dx.doi.org/10.17605/OSF.IO/ZW8NP}.
\end{acknowledgements}

\bibliographystyle{spbasic_updated}      
\bibliography{sample_library}   
\end{document}